\documentclass{aa}
\usepackage{graphicx}
\usepackage{times}
\usepackage{mathptm}
\fontfamily{ptmcm}\selectfont
\begin{document}
%\title{Radio Properties of the Shapley Concentration}
%
%\subtitle{V. The A3571 Cluster Complex}

%\title{Radio emission from the A3571 cluster complex in the Shapley 
%Concentration}

\title{Radio emission from the A3571 cluster complex: 
the final stage of a cluster merger?}

\author{ T. Venturi \inst{1}
	\and S. Bardelli  \inst{2}
	\and M. Zagaria   \inst{1,3}
	\and I. Prandoni  \inst{1,3}
	\and R. Morganti  \inst{4}}
	
\offprints{Venturi T.}
\mail{tventuri@ira.bo.cnr.it}

\institute{Istituto di Radioastronomia, CNR, Via\,Gobetti 101, I--40129, 
Bologna, Italy
\and Osservatorio Astronomico, Via Ranzani 1, I--40126 Bologna, Italy
\and Dipartimento di Astronomia, Universit\`a di Bologna, Via Ranzani 1,
I--40126, Bologna, Italy 
\and NFRA, Postbus 2, A7990 Dwingeloo, The Netherlands}

\date{Received / Accepted }

\titlerunning{The A3571 Cluster Complex}
\authorrunning{Venturi T. et al.}

\abstract{In this paper we report and discuss the results of a radio 
survey in the A3571 cluster complex, a structure
located in the Shapley Concentration core, and formed by the
three clusters A3571, A3572 and A3575.
The survey was carried out simultaneously at 22 cm and 13 cm
with the Australia Telescope Compact Array, and led to the detection
of 124 radio sources at 22 cm. The radio source counts in
this region are in agreement with the background counts.
Among the 36 radio sources with optical counterpart, six have measured 
redshift which places them at the distance of the A3571 cluster complex, 
and nine radio sources have optical counterparts most likely members of 
this cluster complex. All radio galaxies emit at low power level, 
i.e. logP$_{22cm}$ (W Hz$^{-1}$) $\le$ 22.6. A number of them
are likely to be starburst galaxies. 
The radio luminosity function of early type galaxies is in agreement 
with that derived by Ledlow \& Owen (1996) if we restrict our analysis 
to the A3571 cluster. On the basis of the multiwavelength properties of 
the A3571 cluster complex, we propose that it is a very advanced
merger, and explain the radio properties derived from our study in the
light of this hypothesis.
} 
\maketitle

\section{Introduction}

Merging between clusters is thought to be responsible for significant 
changes in the physics of the intracluster medium and in the 
emission properties of the galaxy population, as a consequence of
the enormous energy budget involved ($\sim 10^{60}$ ergs).

The effects of this dynamical process on the hot gas are clear 
and well documented (Sarazin 2000), as well as the connection 
between cluster merging and  peculiar
radio sources such as relics and halos (Feretti \& Giovannini 1996).
In particular, these extended radio sources 
are found to be associated with clusters with some degree
of disturbance, suggesting that merging is 
responsible for the reacceleration of the relativistic electrons
(Feretti \& Giovannini, 2001; Ensslin\& Br\"uggen, 2001; Brunetti et al. 2001;
Buote 2001).
On the other hand it is not yet understood whether cluster--cluster   
collision is able to modify the radio emission properties
of a single galaxy and/or affect the statistics 
of the radio source population.

In order to address this issue, we are carrying on a multiwavelength 
study of the central part of the Shapley Concentration, where an unusually
high number of clusters are merging. This situation originated 
from the high local density, that induced peculiar velocities of
the order of $\sim 1000$ km s$^{-1}$.

Zucca et al. (1993) performed a percolation analysis of the Abell-ACO
clusters in order to find superclusters	and to study the grouping 
characteristics as a function of the overdensity with respect to the
mean density of clusters.  
In particular, they found that at the highest
density contrasts ($>40$), the central part of the Shapley Concentration
is fragmented in three main structures  (``cluster complexes" or 
``groups of clusters").   
One is dominated by the cluster A3558 and could be considered the
core of the supercluster; a second is formed by A3528, A3530 and A3532.
Both complexes present very clear signs of ongoing merging (Hanami et al. 
1999; Bardelli et al. 1996; Ettori et al. 1997; Kull \&B\"ohringer 1999; 
Schindler 1996). The third group of clusters in this region is
dominated by A3571, and includes A3572 and A3575.

We surveyed the A3558 and A3528 cluster regions at 22 cm with 
the Australia Telescope Compact Array (ATCA) to 
the limiting flux density of $\sim 0.8$ mJy (Venturi et al. 2000 and 2001). 
It was found that the A3558 complex is characterised by a significant 
lack of radio sources with respect to the cluster sample of 
Ledlow \& Owen (1996), suggesting the possibility that cluster--cluster
collisions would ``swith off"the existing radio sources,
or inhibit the birth of new ones.
On the contrary, the number of radio sources found in the A3528 complex
is statistically consistent with the one of the control sample.
These two different behaviours can be possibly explained in terms 
of merging age: the A3558 complex is at an advanced stage 
(possibly after the first core-core encounter, Bardelli et al. 1998), while 
the A3528 complex is an early merging, where clusters just started to 
interact (see for example the simulations by Ricker \& Sarazin 2001). 

In this paper we present the radio--optical analysis of the
third group of clusters in the central part of the 
Shapley Concentration, i.e. the A3571--A3572--A3575 chain.
In Section 2 we overview optical and X--ray properties of the A3571
complex; in Section 3 we report on
the radio observations; the radio source sample and the optical
identifications are described in Section 4; 
the radio properties of the A3571 complex are presented in Section 5,
and the radio and optical number counts are derived in Section 6.
Discussion and conclusions are given in Section 7.

We will use H$_o$ = 100 km s$^{-1}$Mpc$^{-1}$ and q$_o$ = 0, which 
leads to a linear scale 1 arcsec = 0.55 kpc at the average redshift
of the cluster complex (z = 0.039).
  
\medskip

\section{The A3571 Complex in the Shapley Concentration}

\begin{figure} [!ht]
\resizebox{\hsize}{!}{\includegraphics{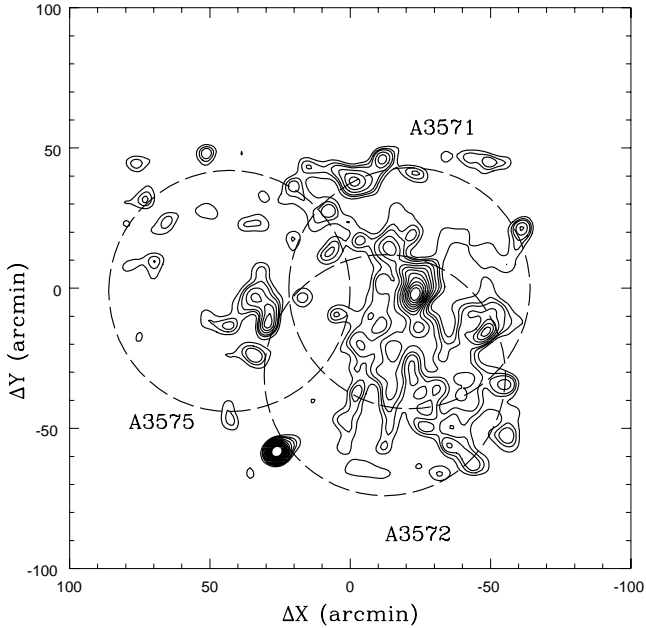}}
\caption{Isocontours of the galaxy density to $b_j$ $=$ 19.5 in the A3571
complex. The figure is centered at $\alpha$ $=$ 13$^h$ 49$^m$ 10$^s$, $\delta$ 
$=$ $-32^{\circ}$ 31$^{\prime}$ $47"$.
Dashed lines correspond to 1 Abell radius circles around cluster
centres, given in Table \ref{opti}. The counts have been smoothed with a 
Gaussian of $6$ arcmin FWHM.}
\label{contour}
\end{figure}

The chain formed by the three clusters A3571--A3572--A3575 
% at $\alpha$ $=$ 13$^h$ 49$^m$ 10$^s$, $\delta$ $=$ $-32^\circ$ 31$^\prime$
% $47"$, 
lies $\sim 5^\circ$ east of the geometrical centre of the Shapley
Concentration, defined by the A3558 complex.
In  Fig. \ref{contour}, we plotted the galaxy isodensity contours
in this region, after smoothing the optical counts (derived from the
COSMOS catalogue) 
with a Gaussian of $6$ arcmin of FWHM. The size of the structure
is  approximately $150\times 70$ arcmin, corresponding to $5.1 \times
2.4$ h$^{-1}$ Mpc at $11550$ km s$^{-1}$ (the average velocity of the
three clusters).
The presence of a significant overdensity at the position
of A3571 is evident, with a tail extending down to the centre of 
A3572. It is difficult to assess if A3572 is part of 
A3571, or if it is an independent structure.
Less clear is the existence of a galaxy excess at the position of A3575.

%%%%%%%%fig 1
%%%%%%%%%%

Table \ref{opti} summarizes the most important features of the three clusters:
in columns 1, 2 and 3 we give respectively name and J2000 coordinates 
of the centre (taken from the ACO catalogue, Abell, Corwin\& Olowin 1989); 
in column 4 the Bautz-Morgan type and richness class; in columns 5 and 6 the 
average recession velocity and the velocity dispersion respectively.

\begin{table}[ht]
%\begin{center}
\caption{Properties of the clusters in the A3571 chain.}
\label{opti}
\begin{tabular}{cccccc}
\hline
 &&&&&\\
 Cluster & RA$_{J2000}$ & DEC$_{J2000}$ & B-M (R)
 & v$_{rec}$ & $\sigma$$_v$ \\
 & $^h$ \hspace{1mm} $^m$ \hspace{1mm} $^s$ 
 &\hspace{1mm} $^\circ$ \hspace{1mm} $^\prime$ &
 & km s$^{-1}$ & km s$^{-1}$ \\
 &&&&&\\
\hline
 &&&&&\\
 A3571 & 13 47 30 & $-32$ 51 &  I (2) & 11730$^a$ & 1022$^a$ \\
 
 A3572 & 13 48 12 & $-33$ 22 &  I-II (0) & 12142$^b$ & 1300$^b$ \\
 
 A3575 & 13 52 36 & $-32$ 52 &  II (0) & 11250$^c$ &  572$^c$  \\
 &&&&&\\
\hline
 &&&&&\\
\end{tabular}
%\end{center}
\\
$^a$ Quintana \& de Souza (1993)\\
$^b$ Drinkwater et al. (1999)\\
$^c$ Quintana et al. (1997)\\
\end{table}

\subsection{A3571} \label{mgc}

The dominant member of the complex is the cluster A3571, with richness class 
$R=2$
(Abell, Corwin \& Olowin, 1989) and Bautz-Morgan type I. 
The mean velocity is $v=11730 \pm 127$ km s$^{-1}$ and the velocity dispersion
is $1022^{+99}_{-77}$ km s$^{-1}$ (Quintana \& de Souza, 1993).
Quintana \& de Souza (1993) suggested that the velocity distribution 
is asymmetric and that the bi-dimensional galaxy distribution
presents some indications of substructure. 

A3571 is characterised by the presence of the large
cD galaxy MCG05$-$33$-$002, located at its centre.
This galaxy, with an extension of $0.1\times 0.3$ h$^{-1}$ Mpc,
is elongated along the north-south direction (Kemp \& Meaburn 1991),
in the same direction of the bi-dimensional galaxy distribution.

As noted by Nevalainen et al. (2000 and 2001), 
while the galaxy distribution indicates some degree of non-relaxation,
the X--ray observations show that A3571 is a well relaxed cluster,
with the presence of a cooling flow.
The hot gas temperature is $7.6$ keV (Nevalainen et al. 2001), consistent
with the galaxy velocity dispersion.
Nevalainen et al. (2001) determined also the total mass using the
hydrostatic equilibrium of the hot gas, and found $M_{tot}\sim 4.6
\times 10^{14}$ h$^{-1}$ M$_{\odot}$. For comparison,
the results of the virial mass of Girardi et al. (1998)
is $M_{tot}\sim 9 \times 10^{14}$ h$^{-1}$ M$_{\odot}$. 

\subsection{A3572 and A3575}
Little is known about the other two clusters in the complex.

A3572 has a richness class R = 0 and is located south of A3571 in the 
plane of the sky (see Fig. \ref{contour}).
Drinkwater et al. (1999) found a mean velocity of 
$<v>=12142$ km s$^{-1}$, which implies an average redshift $<$z$>$ = 0.0405.

A3575 has R = 0 and lies east of A3571. 
Quintana et al. (1997) found $<v>=11250$ km s$^{-1}$ (i.e. $<$z$>$ = 0.0375) 
and $\sigma=572$ km s$^{-1}$ on the basis of only 12 velocities. 

No information is available in the literature on the X--ray properties
of these two clusters. Visual inspection of the ROSAT All Sky 
Survey images does not show any indication of X--ray emission.

\section{Observations and Data Reduction}

\subsection{22 cm and 13 cm ATCA Observations}\label{22cm}

We imaged the A3571 complex with the Australia Telescope
Compact Array (ATCA) at 22 cm ($\nu$ $=$ 1.38 GHz) and 13 cm 
($\nu = 2.38$ GHz) simultaneously. The whole region was covered 
with six different pointings, whose centres are listed in 
Table \ref{obs}. The separation between contiguous fields is 20 arcmin.

%The half-power primary beam at 22 cm is $\sim$ 33$^{\prime}$, while the 
%separation between contiguous field centres is $\sim$ 20$^{\prime}$.
%This allows good overlap between the fields and compensates for the 
%sensitivity loss due to the primary beam attenuation, as clear from
%Fig. \ref{pointings}.
%

\begin{table}[h]
%\begin{center}
\caption{Details on the observations.}
\label{obs}
%\vspace{-5mm}
\begin{tabular}{ccccc}
\hline
&&&&\\
Field & RA$_{J2000}$ & DEC$_{J2000}$ & rms$_{22cm}$  &  rms$_{13cm}$  \\
      & $^h$ \hspace{1mm} $^m$ \hspace{1mm} $^s$ 
      & \hspace{1mm} $^\circ$ \hspace{1mm} $^\prime$
      & mJy b$^{-1}$ & mJy b$^{-1}$ \\
&&&&\\
\hline
&&&&\\
 \#1 & 13 47 30 & $-32$ 51 & 0.12 & 0.12 \\
 \#2 & 13 49 11 & $-32$ 51 & 0.11 & 0.11 \\
 \#3 & 13 50 52 & $-32$ 51 & 0.11 & 0.12 \\
 \#4 & 13 52 31 & $-32$ 51 & 0.12 & 0.12 \\
 \#5 & 13 47 57 & $-33$ 10 & 0.11 & 0.11 \\
 \#6 & 13 48 12 & $-33$ 25 & 0.11 & 0.11 \\
&&&&\\
\hline
\end{tabular}
%\end{center}
\end{table}

%%%%%%%%fig 2
\begin{figure} [!ht]
\resizebox{\hsize}{!}{\includegraphics{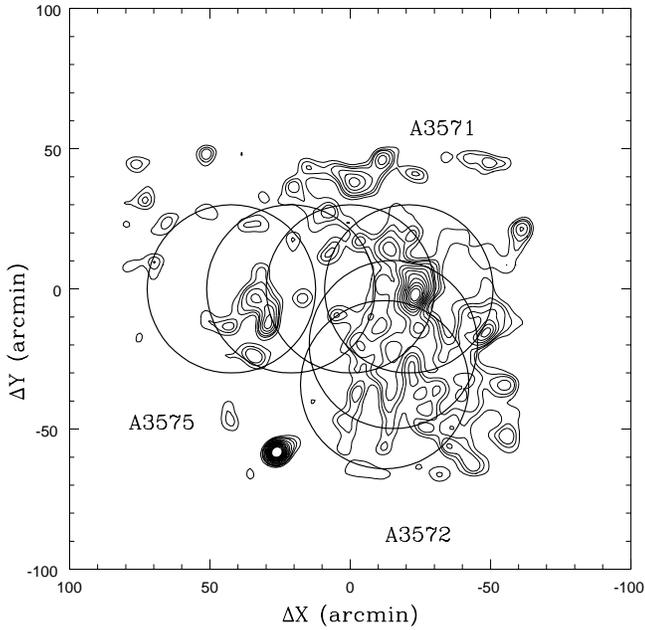}}
\caption{Fields of the observations overlaid on the 
the galaxy density (same as in Fig. 1).
The radius of each circle is 30$^{\prime}$.
The centres of each field are given in Table \ref{obs}.}
\label{pointings}
\end{figure}
%%%%%%%%%%

The observations were carried out in 1995 (December 22 and 23),
using the mosaicing facility of the ATCA.
To obtain a good hour angle coverage, the six fields were observed 
in sequence, for two minutes each. The cycle (including
calibration) was repeated for $2 \times 12$ hours.
The array operated is the 6.0C configuration (full ATCA resolution), 
which yields a resolution of 
$\sim$ 11$^{\prime\prime}$ $\times$ 5$^{\prime\prime}$ at the declination 
of the A3571 complex. The largest detectable structure size 
is $\sim 3$ arcmin.

We observed in continuum mode  with a 128-MHz bandwidth, divided in 
32$~\times~$4 MHz channels, in order to reduce the bandwidth smearing 
effects at large distance from the pointing centres.
B1934$-$638 was used as primary flux density calibrator, 
with assumed flux densities $S_{22cm}~=~14.9$ Jy and 
$S_{13cm}~=~11.6$ Jy.

\subsection{Data Reduction and Image Analysis}

The data reduction was carried out with the package $MIRIAD$ (Sault,
Teuben \& Wright, 1995), which is particular suited for the ATCA observations.
Multifrequency synthesis techniques are implemented, and this allows proper 
gridding of the data in order to reduce
bandwidth smearing effects. This is an important feature, given that we want
to image the fields as large as possible. 
We reduced the data and imaged each field separately.
The six cleaned 22 cm images were then mosaiced (i.e., linearly combined) 
using the $MIRIAD$ task $LINMOS$, in order to obtain a uniform sensitivity
over the whole region. The area covered by our mosaic is $\sim$ 2.4 deg$^2$.
The image analysis was carried out with the $AIPS$ package.

As shown in Fig. \ref{pointings}, a pointing separation of 20 arcmin 
allows a good overlap between fields at 22 cm, where the primary
beam size is $FWHM \sim 33^{\prime}$, and it is best suited to get uniform
sensitivity over the central region of the radio mosaic (Prandoni et 
al. 2000a). On the other hand, the primary beam size at 13 cm
($FWHM \sim 22^{\prime}$) is comparable to the distance between
adjacent fields, and the very little overlap does not allow to compensate
for the radial sensitivity loss due to the primary beam attenuation.
The mosaic technique is not effective at this frequency, therefore in
this case we treated each image separately.

The average rms noise values measured in the individual fields at 
both 22 cm and 13 cm are listed in Table \ref{obs}; the average
rms flux density measured over the whole mosaic is 
$\sigma~=$ 0.13 mJy b$^{-1}$.
At both wavelengths we assumed a source detection flux limit 
S = 0.65 mJy b$^{-1}$ (i.e. $\sim 5\sigma$).
Such limit translates into a radio power detection limit 
logP$_{22}$ (W HZ${-1}$) = 21.02 at the average distance of the A3571 
cluster complex.

\medskip
The internal uncertainty of the source flux density and position
measures can be estimated following Prandoni et al. (2000b),
since our ATCA interferometric data are very similar 
to those presented in their work (same frequency and array configuration).

For unresolved (pointlike) radio sources the uncertainty 
$\Delta$S, associated with the flux density depends on the flux density 
itself ($S_{peak}$), on the local signal-to-noise 
ratio, defined as $S_{peak}/\sigma$, and on the calibration residual errors, 
as follows: 

\begin{equation}
\Delta S = \sigma(S_{peak})/S_{peak} = 0.93 \left( \frac{S_{peak}}{\sigma} \right)^{-1}
\hspace{1mm}.
\end{equation}

Using this equation, a radio source with peak flux density $S_{peak}=$ 0.65 
mJy b$^{-1}$ and signal-to-noise ratio of 5, has $\Delta S~=~$ 0.12 mJy.

The internal position error of the radio sources depends again on
the local signal-to-noise ratio, and on the synthesized beam,
according to the following formulas (Prandoni et al. 2000b):

\begin{equation}
\sigma_\alpha \simeq \frac{b_{min}}{2} \left( \frac{S_{peak}}{\sigma}
\right)^{-1} \hspace{1mm},
\end{equation}

\begin{equation}
\sigma_\delta \simeq \frac{b_{maj}}{2} \left( \frac{S_{peak}}{\sigma}
\right)^{-1} \hspace{1mm},
\end{equation}

where we assumed $b_{min}$ $=$ 5$^{\prime\prime}$ and  $b_{maj}$ $=$
11$^{\prime\prime}$, the average synthesized beam values of the 22 cm survey.
Radial deformations due to the cromatic aberration are negligible in our 
mosaic, so we did not compensate for this effect.
We assumed for all  sources a conservative position uncertainty of 
1$^{\prime\prime}$, both in RA and in DEC, which corresponds to the 
values for the weakest sources in the sample, i.e. $5\sigma$ detections.

\section{Results}
\subsection{The Sample of Radio Sources}

We detected a total of 124 sources at 22 cm above 0.65 mJy in
the mosaic image. We point out that due to the smaller field of view 
of the 13 cm observations, and to possible spectral index effects, 
only a fraction of the sources detected at 22 cm was revealed at this 
shorter wavelength. In particular at 13 cm we detected 36/124 sources, 
i.e. 29\%.

The source list is reported in Table \ref{catal1},
where we give name and J2000 position (columns 1, 2 and 3); integrated 
flux  density at 22 cm corrected for the primary beam attenuation 
(column 4); integrated flux  density at 13 cm corrected for the primary 
beam attenuation  (column 5); spectral index $\alpha_{13}^{22}$ for those 
sources detected at both frequencies (column 6), to be read in the
sense S$~\propto~\nu^{-\alpha}$; radio morphology (column 7). 
For those sources undetected at 13 cm, a $\star$ is reported in Table 
\ref{catal1}. 
%We note that this flux density upper limit is not corrected
%for the primary beam attenuation.

By inspecting our 22 cm mosaic image, we classified the radio morphology of the 
sources as follows: unresolved (unres.), resolved/extended (res.) 
and double (D). For the 
extended sources in the sample we give the position of the radio peak, 
while for the two double systems we give the coordinates of the barycentre.
The majority of the radio sources, 110 out of 124 ($\sim$89\%), are 
unresolved at the resolution of the final image, 2 ($\sim$2\%) have a 
double structure and the remaining 12 ($\sim$9\%) are extended. 

\begin{figure} [!ht]
\resizebox{\hsize}{!}{\includegraphics{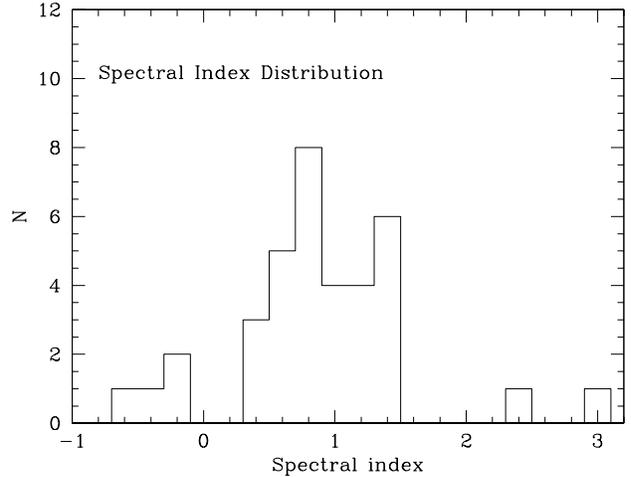}}
\caption{Distribution of the spectral index $\alpha_{13}^{22}$ for those
sources in the sample detected at both wavelengths.}
\label{alfa}
\end{figure}

In Fig. \ref{alfa} we show the distribution of the spectral index
for those sources detected at both 22 cm and 13 cm. We point out
that the computation of the spectral index was done using the 
total flux density as measured on the images, so the
different $u-v$ coverage at the two wavelengths was not taken into
account. For extended/resolved radio sources, this may have biassed our 
analysis towards spectra steeper than their intrinsic shape, given the 
lack of short spacings at 13 cm.
This effect is particularly evident for the two sources 
with the steepest spectra: they are both extended sources (see Table 
\ref{catal1}), and we believe that a fraction of their extended flux 
was missed at 13 cm.
From Fig. \ref{alfa} we conclude that the sample of radio sources
includes both compact active nuclei with flat and inverted spectrum, and
radio sources with normal to moderately steep spectrum.

%% figure 3
%

\begin{table*}[t]
%\begin{center}
\setcounter{table}{2}
\caption{Source list and flux density values.}
  \label{catal1}
\begin{tabular}{lllrrrc}
\hline
&&&&&&\\
 Name  & RA$_{J2000}$  & DEC$_{J2000}$ & S$_{22cm}$& S$_{13cm}$ & 
$\alpha_{13}^{22}$ & Morph. \\
       & $^h$ \hspace{1mm} $^m$ \hspace{1mm} $^s$ 
& $^\circ$ \hspace{1mm} $^\prime$ \hspace{1mm} $^{\prime\prime}$ &
  mJy  &   mJy   &&	   \\
&&&&&&\\
\hline
&&&&&&\\
J1346$-$3259   &  13 46 13.0 &  $-$32 59 33  &	2.41  &$\star$ &&  unres.\\
J1346$-$3252   &  13 46 13.7 &  $-$32 52 28  &	3.90  &$\star$ &&  unres.\\
J1346$-$3231   &  13 46 25.9 &  $-$32 31 39  &	5.57  &$-$     &&  unres.\\
J1346$-$3237   &  13 46 34.3 &  $-$32 37 08  &	1.89  &$\star$ &&  unres.\\
J1346$-$3235   &  13 46 38.8 &  $-$32 35 55  & 12.59  &$\star$ &&  unres.\\
J1346$-$3303   &  13 46 39.4 &  $-$33 03 44  &	1.04  &$\star$ &&  unres.\\
J1346$-$3254   &  13 46 40.9 &  $-$32 54 23  &	1.20  &$\star$ &&  unres.\\
J1346$-$3305   &  13 46 47.6 &  $-$33 05 03  &	4.45  &$\star$ &&  unres.\\
J1346$-$3320   &  13 46 50.3 &  $-$33 20 15  &	1.09  &$\star$ &&  unres.\\
J1346$-$3304   &  13 46 50.4 &  $-$33 04 59  &	4.20  &$\star$ &&  unres.\\
J1346$-$3248   &  13 46 53.4 &  $-$32 48 22  &	1.69  &$\star$ &&  unres.\\
J1346$-$3346   &  13 46 59.0 &  $-$33 46 16  &	4.47  &$-$     &&  unres.\\
J1347$-$3311a  &  13 47 02.7 &  $-$33 11 51  &	0.89  &$\star$ &&  unres.\\
J1347$-$3332   &  13 47 07.3 &  $-$33 32 06  &	2.32  &$\star$ &&   res. \\
J1347$-$3339a  &  13 47 08.1 &  $-$33 39 45  &	5.52  &$-$     &&  unres.\\
J1347$-$3306   &  13 47 09.0 &  $-$33 06 46  &	0.84  &$\star$ &&  unres.\\
J1347$-$3313   &  13 47 09.4 &  $-$33 13 51  &	1.35  &$\star$ &&  unres.\\
J1347$-$3334   &  13 47 20.2 &  $-$33 34 47  &	2.29  &$\star$ &&  unres.\\
J1347$-$3252$^{\diamondsuit}$
               & 13 47 24.6 &  $-$32 52 15  & 24.23  &  11.23 & 1.41&   D   \\
J1347$-$3325   &  13 47 24.2 &  $-$33 25 11  & 16.29  &   4.55 & 2.34&  res. \\
J1347$-$3256   &  13 47 25.3 &  $-$32 56 10  &	1.34  &$\star$ &&  unres.\\
J1347$-$3324   &  13 47 25.4 &  $-$33 24 41  &	7.08  &$\star$ &&   res. \\
J1347$-$3229   &  13 47 26.5 &  $-$32 29 30  & 54.39  & -      &&  unres.\\
J1347$-$3314   &  13 47 26.8 &  $-$33 14 33  &	1.04  &$\star$ &&  unres.\\
J1347$-$3251   &  13 47 28.5 &  $-$32 51 54  &  4.05  &   2.52 & 0.87& unres.\\
J1347$-$3310   &  13 47 31.4 &  $-$33 10 04  &	3.83  &   2.80 & 0.57& unres.\\
J1347$-$3241   &  13 47 34.3 &  $-$32 41 31  &	2.96  &   1.54 & 1.20& unres.\\
J1347$-$3323   &  13 47 36.3 &  $-$33 23 28  &	3.38  &   1.92 & 1.04& unres.\\
J1347$-$3339b  &  13 47 37.3 &  $-$33 39 27  &	7.13  & 7.76 &$-$0.16& unres.\\
J1347$-$3311b  &  13 47 42.4 &  $-$33 11 43  & 15.60  &   9.15 & 0.98& unres.\\
J1347$-$3328   &  13 47 48.7 &  $-$33 28 40  &	4.12  &   1.90 & 1.42& unres.\\
J1347$-$3259   &  13 47 53.1 &  $-$32 59 01  &	1.08  &$\star$ &&  unres.\\
J1347$-$3236   &  13 47 54.2 &  $-$32 36 59  & 14.18  &  10.44& 0.56& unres.\\
J1347$-$3301   &  13 47 56.8 &  $-$33 01 06  &	9.09  &   6.12 & 0.73& unres.\\
J1347$-$3330$^{\clubsuit}$
               &  13 47 57.5 &  $-$33 30 02  &	5.06  &   3.79 & 0.53& unres.\\
J1347$-$3302   &  13 47 59.5 &  $-$33 02 03  &	0.98  &$\star$ &&  unres.\\
J1348$-$3311   &  13 48 04.2 &  $-$33 11 33  &	1.13  &   0.74 & 0.78& unres.\\
J1348$-$3315   &  13 48 04.9 &  $-$33 15 58  &	1.47  &$\star$ &&  unres.\\
J1348$-$3246   &  13 48 08   &  $-$32 46 15  & 192.08 & 102.54 & 1.15&   D   \\
J1348$-$3339   &  13 48 10.5 &  $-$33 39 31  &	2.25  &$\star$ &&  unres.\\
J1348$-$3323   &  13 48 14.3 &  $-$33 23 32  &	2.22  &$\star$ &&  unres.\\
J1348$-$3239   &  13 48 15.3 &  $-$32 39 31  &	1.04  &$\star$ &&  unres.\\
J1348$-$3255   &  13 48 16.1 &  $-$32 55 33  &	0.75  &$\star$ &&  unres.\\
J1348$-$3234   &  13 48 24.9 &  $-$32 34 35  &	4.06  &$\star$ &&  unres.\\
J1348$-$3249   &  13 48 30.3 &  $-$32 49 10  &	1.03  &$\star$ &&  unres.\\
J1348$-$3256   &  13 48 32.3 &  $-$32 56 33  &	1.82  &$\star$ &&  unres.\\
J1348$-$3318   &  13 48 33.5 &  $-$33 18 47  &	0.92  &$\star$ &&  unres.\\
J1348$-$3328   &  13 48 35.1 &  $-$33 28 18  &	0.69  &$\star$ &&  unres.\\
J1348$-$3310   &  13 48 37.5 &  $-$33 10 04  & 11.70  &   6.75 & 1.01& unres.\\
J1348$-$3258   &  13 48 39.6 &  $-$32 58 16  &	1.22  & 1.46 &$-$0.33& unres.\\
&&&&&&\\
\hline
\end{tabular}
%\end{center}
%\vspace{5mm}
\end{table*}

\begin{table*}[h]
\setcounter{table}{2}
\caption{$-$ continued}
%\begin{center}
\begin{tabular}{lllrrrc}
\hline
&&&&&&\\
 Name	 & $RA_{J2000}$  & $DEC_{J2000}$  & S$_{22cm}$& S$_{13cm}$ & 
$\alpha_{13}^{22}$ & Morph. \\
	     & $^h$ \hspace{1mm} $^m$ \hspace{1mm} $^s$ 
& $^\circ$ \hspace{1mm} $^\prime$ \hspace{1mm} $^{\prime\prime}$&
mJy   &   mJy   &&	   \\
&&&&&&\\
\hline
&&&&&&\\
J1348$-$3257   &  13 48 46.3 &  $-$32 57 37  &	6.42  &   4.04 & 0.85& unres.\\
J1348$-$3241   &  13 48 46.5 &  $-$32 41 41  &	4.35  &   3.37 & 0.47&  res. \\
J1348$-$3303   &  13 48 47.7 &  $-$33 03 16  & 14.15  &   6.52 & 1.42& unres.\\
J1348$-$3305   &  13 48 52.9 &  $-$33 05 17  &	3.22  &$\star$ &&  unres.\\
J1348$-$3304   &  13 48 55.6 &  $-$33 04 59  &	3.74  &$\star$ &&  unres.\\
J1348$-$3250   &  13 48 58.9 &  $-$32 50 45  &	0.76  &$\star$ &&  unres.\\
J1348$-$3337   &  13 48 59.8 &  $-$33 37 55  &	2.68  &$\star$ &&  unres.\\
J1349$-$3326   &  13 49 02.3 &  $-$33 26 57  &	0.93  &$\star$ &&  unres.\\
J1349$-$3243   &  13 49 07.8 &  $-$32 43 36  &	3.00  &$\star$ &&   res. \\
J1349$-$3320a  &  13 49 10.8 &  $-$33 20 12  &	3.70  &$\star$ &&  unres.\\
J1349$-$3317   &  13 49 15.0 &  $-$33 17 50  &	2.82  &$\star$ &&  unres.\\
J1349$-$3304   &  13 49 15.6 &  $-$33 04 43  &	3.27  &   2.39 & 0.58& unres.\\
J1349$-$3247   &  13 49 18.8 &  $-$32 47 28  &	1.93  &   0.89 & 1.42& unres.\\
J1349$-$3314a  &  13 49 23.9 &  $-$33 14 08  &	5.83  &$\star$ &&  unres.\\
J1349$-$3323   &  13 49 26.0 &  $-$33 23 56  & 23.95  &$\star$ &&  unres.\\
J1349$-$3329   &  13 49 27.0 &  $-$33 29 27  &	0.73  &$\star$ &&  unres.\\
J1349$-$3246   &  13 49 27.7 &  $-$32 46 11  &	1.13  &$\star$ &&  unres.\\
J1349$-$3230   &  13 49 27.9 &  $-$32 30 10  &	1.72  &$-$     &&  unres.\\
J1349$-$3238a  &  13 49 31.5 &  $-$32 38 06  &	1.59  &$-$     &&  unres.\\
J1349$-$3351   &  13 49 35.1 &  $-$33 51 04  &212.79  &$-$     &&  unres.\\
J1349$-$3314b  &  13 49 36.1 &  $-$33 14 34  & 10.28  &$-$     &&  unres.\\
J1349$-$3301   &  13 49 40.5 &  $-$33 01 19  &	2.66  &$\star$ &&  unres.\\
J1349$-$3314c  &  13 49 40.6 &  $-$33 14 28  &	9.32  &$-$     &&  unres.\\
J1349$-$3244   &  13 49 40.8 &  $-$32 44 38  &	1.92  &$\star$ &&  unres.\\
J1349$-$3320b  &  13 49 43.1 &  $-$33 20 18  & 14.86  &$-$     &&  unres.\\
J1349$-$3238b  &  13 49 58.2 &  $-$32 38 49  &	5.56  & 7.35 &$-$0.51& unres.\\
J1349$-$3245   &  13 49 58.9 &  $-$32 45 04  &	7.02  &   4.55 & 0.80& unres.\\
J1350$-$3334   &  13 50 01.2 &  $-$33 34 22  & 10.00  &$-$     &&  unres.\\
J1350$-$3302   &  13 50 08.8 &  $-$33 02 19  &	4.37  &$\star$ &&   res. \\
J1350$-$3311   &  13 50 09.1 &  $-$33 11 24  &	6.01  &$-$     &&  unres.\\
J1350$-$3301   &  13 50 12.1 &  $-$33 01 58  &	1.21  &$\star$ &&  unres.\\
J1350$-$3249   &  13 50 18.7 &  $-$32 49 17  &	2.55  &   1.66 & 0.79& unres.\\
J1350$-$3336   &  13 50 22.5 &  $-$33 36 58  & 56.99  &$-$     &&  unres.\\
J1350$-$3243   &  13 50 26.8 &  $-$32 43 38  &  9.22  &   6.16 & 0.74& unres.\\
J1350$-$3248a  &  13 50 46.3 &  $-$32 48 51  &	1.35  &$\star$ &&  unres.\\
J1350$-$3250   &  13 50 51.7 &  $-$32 50 17  &	1.08  &   0.91 & 0.31& unres.\\
J1350$-$3248b  &  13 50 57.1 &  $-$32 48 48  &	1.07  &$\star$ &&  unres.\\
J1351$-$3248   &  13 51 24.6 &  $-$32 48 49  &	2.06  &$\star$ &&  unres.\\
J1351$-$3238   &  13 51 28.4 &  $-$32 38 35  &  0.91  &$\star$ &&  unres.\\
J1351$-$3259a  &  13 51 29.4 &  $-$32 59 11  &	1.50  &$\star$ &&  unres.\\
J1351$-$3249   &  13 51 33.8 &  $-$32 49 49  &	2.80  &   1.31 & 1.39& unres.\\
J1351$-$3228a  &  13 51 34.1 &  $-$32 28 44  & 32.75  &$-$     &&  unres.\\
J1351$-$3228b  &  13 51 35.5 &  $-$32 28 45  & 57.82  &$-$     &&  unres.\\
J1351$-$3234   &  13 51 35.5 &  $-$32 34 13  &  3.42  &$-$     &&  unres.\\
J1351$-$3247   &  13 51 35.9 &  $-$32 47 10  &	3.14  &$\star$ &&   res. \\
J1351$-$3259b  &  13 51 41.6 &  $-$32 59 11  &	3.29  &$\star$ &&  unres.\\
J1351$-$3313   &  13 51 43.8 &  $-$33 13 48  &	1.67  &$-$     &&  unres.\\
J1351$-$3303   &  13 51 44.0 &  $-$33 03 07  &	0.94  &$\star$ &&  unres.\\
J1351$-$3310   &  13 51 46.2 &  $-$33 10 42  &	2.79  &$-$     &&  unres.\\
J1351$-$3301   &  13 51 56.9 &  $-$33 01 46  &	1.60  &$\star$ &&  unres.\\
&&&&&&\\
\hline
\end{tabular}
%\end{center}
%\vspace{5mm}
\label{tab:catal2}
\end{table*}

\begin{table*}[h]
\setcounter{table}{2}
\caption{$-$ continued}
%\begin{center}
\begin{tabular}{lllrrrc}
\hline
&&&&&&\\
 Name	     & $RA_{J2000}$  & $DEC_{J2000}$  & S$_{22cm}$& S$_{13cm}$ &
$\alpha_{13}^{22}$ & Morph. \\
	     & $^h$ \hspace{1mm} $^m$ \hspace{1mm} $^s$ 
& $^\circ$ \hspace{1mm} $^\prime$ \hspace{1mm} $^{\prime\prime}$&
mJy   &   mJy   &&	   \\
&&&&&&\\
\hline
&&&&&&\\
J1351$-$3240   &  13 51 59.8 &  $-$32 40 19  &	2.02  &$\star$ &&  unres.\\
J1351$-$3229   &  13 51 59.9 &  $-$32 29 24  &	2.94  &$-$     &&  unres.\\
J1352$-$3314   &  13 52 08.0 &  $-$33 14 45  & 20.42  &$-$     &&   res. \\
J1352$-$3315   &  13 52 09.1 &  $-$33 15 26  & 26.16  &$-$     &&   res. \\
J1352$-$3236   &  13 52 14.2 &  $-$32 36 44  &	1.22  &$\star$ &&  unres.\\
J1352$-$3243a  &  13 52 16.5 &  $-$32 43 38  &	1.24  &$\star$ &&  unres.\\
J1352$-$3245   &  13 52 24.8 &  $-$32 45 37  & 12.42  &   6.49 & 1.19&  res. \\
J1352$-$3308   &  13 52 32.9 &  $-$33 08 49  &  2.37  &$\star$ &&  unres.\\
J1352$-$3306   &  13 52 33.1 &  $-$33 06 01  &	6.52  &   2.95 & 1.46& unres.\\
J1352$-$3243b  &  13 52 34.1 &  $-$32 43 32  &	6.56  &   3.51 & 1.15& unres.\\
J1352$-$3244   &  13 52 34.9 &  $-$32 44 33  & 11.00  &   6.13 & 1.07& unres.\\
J1352$-$3248   &  13 52 40.2 &  $-$32 48 56  &	1.34  &   1.07 & 0.41& unres.\\
J1352$-$3230a  &  13 52 42.6 &  $-$32 30 51  &	5.23  &$-$     &&  unres.\\
J1352$-$3302   &  13 52 43.5 &  $-$33 02 29  &	8.13  & 8.92 &$-$0.17& unres.\\
J1352$-$3230b  &  13 52 46.3 &  $-$32 30 24  & 10.22  &$-$     &&  unres.\\
J1352$-$3301   &  13 52 50.2 &  $-$33 01 58  & 16.70  &   3.32 & 2.96&  res. \\
J1352$-$3251   &  13 52 50.5 &  $-$32 51 24  &232.18  & 145.41 & 0.86& unres.\\
J1352$-$3304   &  13 52 56.6 &  $-$33 04 40  & 37.85  &  27.57 & 0.58& unres.\\
J1353$-$3229   &  13 53 08.4 &  $-$32 29 27  &	5.81  &$-$     &&  unres.\\
J1353$-$3228   &  13 53 20.1 &  $-$32 28 03  &	6.00  &$-$     &&  unres.\\
J1353$-$3238   &  13 53 22.5 &  $-$32 38 11  &	1.44  &$\star$ &&  unres.\\
J1353$-$3239   &  13 53 27.2 &  $-$32 39 17  &	4.68  &$\star$ &&  unres.\\
J1353$-$3300   &  13 53 45.5 &  $-$33 00 01  &	4.18  &$\star$ &&  unres.\\
J1354$-$3251   &  13 54 24.1 &  $-$32 51 02  & 14.66  &$-$     &&   res. \\
&&&&&\\
\hline
\end{tabular}
%\end{center}
\vspace{1mm}

Notes to Table 3.

$\star$ Non detections at 13 cm, i.e, S$_{13} \le$ 0.65 mJy in the image
not corrected for the primary beam attenuation.

$^{\diamondsuit}$ The position of this radio source was taken from the 13 
cm image, where the nucleus of the double structure is clearly visible.

$^{\clubsuit}$ This source shows a double morphology in the 13 cm image.

\label{tab:catal3}
\end{table*}

\subsection{Optical Identifications}

We searched for optical counterparts of the radio sources detected 
in our 22 cm survey using two different catalogues: the COSMOS/UKST
Southern Sky Object Catalogue (Yentis et al., 1992), limited to $b_j$ $=$ 19.5,
and the APM Catalogue (Automated Photographic Measuring), limited to
$b_j$ $=$ 20 (Maddox et al. 1990).
Both catalogues are the result of automatic scans of the UK Schimdt Telescope
(UKST) plates and have a claimed positional accuracy of
$\sim$0.25$^{\prime\prime}$. The transformation from the plate frame to the
sky, however, introduces further uncertainties, so we adopted a 
conservative mean positional error of
1.5$^{\prime\prime}$ (Unewisse et al. 1993).

Due to incompleteness effects at both faint and bright magnitudes 
(bright galaxies can be resolved out) all radio sources were 
overplotted on the Digitized Sky Survey (DSS) images and scrutinized by eye, 
to make sure that no identification was missed.

Given the uncertainties in the radio and optical positions, we estimated the
reliability of our identifications using the parameter $R$, defined as:

$$
R^2 = \frac{\Delta^2_{r-o}}{\sigma^2_g + \sigma^2_r}\hspace{1mm}
$$
where $\Delta_{r-o}$ is the positional offset between the radio and
optical coordinates, $\sigma_g$ is the optical position error and 
$\sigma_r$ is the uncertainty in the radio position.

For point-like radio sources and point-like optical counterparts, we 
considered reliable identifications all matches with $R~\leq~3$; 
for the extended radio sources and/or extended optical galaxies, the value 
of $R$ is less meaningful and we trusted the direct eye inspection.

Finally, we used NED ($NASA$ $Extragalactic$ $Database$) to search 
for the redshift information of the optical counterparts, and considered
as belonging to the Shapley Concentration all galaxies with recession
velocities in the range 10000 $<~v_r$ (km s$^{-1}$) $<$ 18000.\\

We found 36 optical counterparts, $\sim$ 29\% of the total, 6 of
which with known redshift and all belonging to the A3571 complex ($\sim$ 17\% 
of the 36 sources identified). The list of the radio--optical identifications 
is given in Table \ref{opt}, which is organised
as follows: 
column 1 = radio and optical name (where available);  
columns 2 and 3 = coordinates (J2000) of the radio source and of the 
optical counterpart; 
column 4 = flux density of the radio source at 22 cm and the $b_j$ magnitude 
of the counterpart;
column 5 = flux density at 13 cm;
column 6 = spectral index $\alpha^{22}_{13}$ and optical
catalogue used to retrieve the information (A = APM, C = COSMOS);
column 7 = monochromatic radio power at 22 cm for the
radio galaxies with known redshift and absolute magnitude $B_j$ of the
optical counterpart; 
column 8 = radio morphology and parameter $R$;
column 9 = cluster where the radio galaxy is located and recession velocity.

In a number of cases the large optical extent of the galaxy and/or
the extent of the radio emission lead to $R~>~$ 3. For these sources we
considered the identification as reliable if the optical counterpart falls 
within the radio isophotes. A note to Table \ref{opt} clarifies these cases.

\begin{figure} [!ht]
%\hspace{3cm}\resizebox{12cm}{!}{\includegraphics{sandrorad.ps}}
\resizebox{\hsize}{!}{\includegraphics{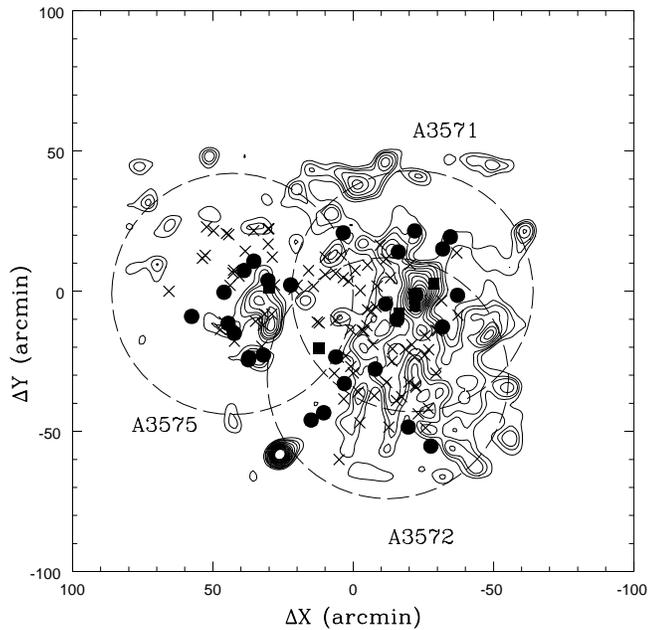}}
\caption{Location of the radio sources overlaid on the optical isodensities of
the A3571 complex. Filled squares represent the optically identified
radio galaxies with measured
redshift, filled circles those without redshift information and crosses stand
for the radio sources without optical counterparts.}
\label{sovr}
\end{figure}

The location of the radio sources overlaid on the optical isodensities of the
A3571 complex is shown in Fig. \ref{sovr}. We note that the radio sources
are distributed fairly uniformly over the region covered by our
observations, regardless of the underlying optical density, which
varies considerably going from A3571 to the other clusters in the
chain. From Table \ref{opt} it is clear
that the Shapley radio galaxies, i.e. galaxies with measured redshift,
are located in A3571 (4/6) and in
the region between the centre of A3571 and A3575 (2/6).

%% figure 4

%\clearpage

\begin{table*}[t]
\setcounter{table}{3}
%\begin{center}
\caption{Optical identifications.}
  \label{opt}
\begin{tabular}{lllrrcrcc}
\hline
&&&&&&&&\\

Radio  & RA$_{J2000}$ & DEC$_{J2000}$ & S$_{22cm}$ & S$_{13cm}$&
$\alpha^{22}_{13}$ & logP$_{22cm}$ & Radio Type   & Cluster \\

& $^h$ \hspace{1mm} $^m$ \hspace{1mm} $^s$ &
$^\circ$ \hspace{1mm} $^\prime$ \hspace{1mm} $^{\prime\prime}$		   
&  mJy	& mJy  & &W Hz$^{-1}$  & &     \\

Optical  & RA$_{J2000}$ & DEC$_{J2000}$ & $b_j$	& &Cat.
& $B_j$ & $R$  &  v$_r$ \\
& $^h$ \hspace{1mm} $^m$ \hspace{1mm} $^s$	   &
$^\circ$ \hspace{1mm} $^\prime$ \hspace{1mm} $^{\prime\prime}$
&&&&&&km s$^{-1}$ \\

&&&&&&&&\\
\hline
&&&&&&&&\\
J1346$-$3252 & 13 46 13.7 & $-$32 52 28 &  3.90 &&   && unres. &   \\
	     & 13 46 13.8 & $-$32 52 26 & 21.80 && A && 0.90   &   \\
&&&&&&&&\\
J1346$-$3231 & 13 46 25.9 & $-$32 31 39 &  5.57 &&   && unres. &   \\
             & 13 46 25.9 & $-$32 31 39 & 19.40 && C && 0.06   &   \\
&&&&&&&&\\
J1346$-$3237b& 13 46 34.3 & $-$32 37 08 &  1.89 &&   && unres. & A3571 \\
             & 13 46 34.3 & $-$32 37 05 & 18.13 && C && 1.34   &   \\
&&&&&&&&\\
J1346$-$3235 & 13 46 38.8 & $-$32 35 55 & 12.59 &&   &&   res. &   \\
             & 13 46 38.8 & $-$32 35 54 & 20.31 && C && 0.65   &   \\
&&&&&&&&\\
J1346$-$3303 & 13 46 39.4 & $-$33 03 44 &  1.04 &&   && unres. & A3571 \\
	     & 13 46 39.5 & $-$33 03 44 & 17.78 && C && 1.04   &       \\
&&&&&&&&\\
J1346$-$3248 & 13 46 53.4 & $-$32 48 22 &  1.69 &&   &&   res. &   \\
             & 13 46 53.3 & $-$32 48 20 & 19.95 && C && 0.99   &   \\
&&&&&&&&\\
J1346$-$3346 & 13 46 59.0 & $-$33 46 16 &  4.47 &&   && unres. &   \\
	     & 13 46 58.9 & $-$33 46 12 & 12.51$^a$&& C && 1.60&   \\
&&&&&&&&\\
J1347$-$3252 & 13 47 24.6 & $-$32 52 15 & 24.23 & 11.23 & 1.41 &&  D     &   \\
             & 13 47 24.3 & $-$32 52 13 & 21.98?&& C &&1.00$^b$&   \\
&&&&&&&&\\
J1347$-$3256 & 13 47 25.3 & $-$32 56 10&  1.34 &&   &     21.35&unres. &A3571\\
             & 13 47 25.3 & $-$32 56 13& 15.78 && C &$-$19.59  & 1.57  &11850\\
&&&&&&&&\\
J1347$-$3229 & 13 47 26.5 & $-$32 29 30 & 54.39 &&   &&  res. &    \\
	     & 13 47 26.5 & $-$32 29 30 & 20.42 && C && 0.08  &    \\
&&&&&&&&\\
J1347$-$3251 & 13 47 28.5 & $-$32 51 54&  4.05 &2.52&0.87&21.82&unres.& A3571\\
	     & 13 47 28.3 & $-$32 51 54& 13.81 &&C& $-$21.53   & 1.00 & 11669\\
&&&&&&&&\\
J1347$-$3339 & 13 47 37.3 & $-$33 39 27 &  7.13 & 7.76&$-$0.16 &&  res.  &   \\
	     & 13 47 37.5 & $-$33 39 30 & 22.46 &&   && 1.96   &   \\
&&&&&&&&\\
J1347$-$3259 & 13 47 53.1 & $-$32 59 01&  1.08 &&    &21.20& unres.&A3571 \\
             & 13 47 53.1 & $-$32 59 02& 15.18 &&C&$-$20.03& 0.32 &11100 \\
&&&&&&&&\\
J1347$-$3236 & 13 47 54.2 & $-$32 36 59 & 14.18 & 10.44& 0.56&& unres. &A3571\\
             & 13 47 54.2 & $-$32 36 35 & 19.77 && C && 0.08   &   \\
&&&&&&&&\\
J1347$-$3301 & 13 47 56.8 & $-$33 01 06 &  9.09 & 6.12& 0.73   && unres. &   \\
	     & 13 47 56.7 & $-$33 01 09 & 17.98 && A && 1.33   &   \\
&&&&&&&&\\
J1348$-$3255 & 13 48 16.1 & $-$32 55 33 &  0.75 &&   && unres. &   \\
	     & 13 48 16.2 & $-$32 55 33 & 21.77 && A && 0.20   &   \\
&&&&&&&&\\
\hline
\end{tabular}
%\end{center}
%\vspace{5mm}
\end{table*}

\begin{table*}[t]
%\begin{center}
\setcounter{table}{3}
\caption{$-$ continued}
  \label{opt2}
\begin{tabular}{lllrrcrcc}
\hline
&&&&&&&&\\

 Radio   & RA$_{J2000}$ & DEC$_{J2000}$ & S$_{22cm}$ & S$_{13cm}$&
 $\alpha^{22}_{13}$ & logP$_{22cm}$ & Radio Type   & Cluster \\

	    &	$^h$ \hspace{1mm} $^m$ \hspace{1mm} $^s$	   &
	    $^\circ$ \hspace{1mm} $^\prime$ \hspace{1mm} $^{\prime\prime}$		   
	    &  mJy	& mJy  & &W Hz$^{-1}$
	     & &     \\

 Optical  & RA$_{J2000}$ & DEC$_{J2000}$ & $b_j$	& &Cat.
 & $B_j$ & $R$  &  v$_r$ \\
	    &	$^h$ \hspace{1mm} $^m$ \hspace{1mm} $^s$	   &
	    $^\circ$ \hspace{1mm} $^\prime$ \hspace{1mm} $^{\prime\prime}$
	    &&&&&&km s$^{-1}$ \\

&&&&&&&&\\
\hline
&&&&&&&&\\
J1348$-$3318 & 13 48 33.5 & $-$33 18 47 &  0.92 &&   && unres. &   \\
             & 13 48 33.5 & $-$33 18 48 & 18.57 && C && 0.60   &   \\
&&&&&&&&\\
J1349$-$3243 & 13 49 07.8 & $-$32 43 36&  3.00 &&        &21.83&  res.&A3571\\
             & 13 49 07.7 & $-$32 43 34& 15.24 &&C&$-$20.48    &  1.10&13900\\
&&&&&&&&\\
J1349$-$3323 & 13 49 26.0 & $-$33 23 56 & 23.95 &&   && unres. & A3571/2\\
	     & 13 49 25.9 & $-$33 23 49 & 15.74 && C && 3.45$^c$&  \\
&&&&&&&&\\
J1349$-$3230 & 13 49 27.9 & $-$32 30 10 &  1.72 &&   && unres. & A3571/5 \\
	     & 13 49 27.7 & $-$32 30 11 & 16.33 && C &&  0.93  &   \\
&&&&&&&&\\
J1349$-$3314c& 13 49 40.6 & $-$33 14 28 &  9.32 &&   &&  res.  &   \\
	     & 13 49 40.6 & $-$33 14 28 & 19.40 && C && 0.30   &   \\
&&&&&&&&\\
J1350$-$3334 & 13 50 01.2 & $-$33 34 22 & 10.00 &&   &&  res.  &   \\
             & 13 50 01.3 & $-$33 34 29 & 20.06 && C && 3.22$^c$&  \\  
&&&&&&&&\\
J1350$-$3311 & 13 50 09.1 & $-$33 11 24&  6.01 &&    &22.35&unres.& A3575 \\
             & 13 50 09.1 & $-$33 11 23& 17.43 &&C& $-$18.80 &  0.79& 17688 \\
&&&&&&&&\\
J1350$-$3336 & 13 50 22.5 & $-$33 36 58 & 56.99 &&   && unres. &   \\
	     & 13 50 22.4 & $-$33 36 58 & 21.53 && C && 0.66   &   \\
&&&&&&&&\\
J1350$-$3248b& 13 50 57.1 & $-$32 48 48 &  1.07 &&   && unres. & A3575  \\
	     & 13 50 57.1 & $-$32 48 47 & 18.33 && C && 0.44   &   \\
&&&&&&&&\\
J1351$-$3249 & 13 51 33.8 & $-$32 49 49&  2.80 &1.31&1.39&21.72&unres.& A3575\\
             & 13 51 33.8 & $-$32 49 49& 16.50 &&C&$-$18.99    & 0.06 & 12542\\
&&&&&&&&\\
J1351$-$3247 & 13 51 35.9 & $-$32 47 10&  3.14 &1.34&1.56&     &  res. &  \\
             & 13 51 36.1 & $-$32 47 13& 18.82 && C &&           1.93  &  \\
&&&&&&&&\\
J1351$-$3313 & 13 51 43.8 & $-$33 13 48 &  1.67 &&   && unres. & A3575  \\
             & 13 51 43.7 & $-$33 13 49 & 18.37 && C && 0.85   &   \\
&&&&&&&&\\
J1351$-$3240 & 13 51 59.8 & $-$32 40 19 &  2.02 &&   && unres. & A3575  \\
             & 13 51 59.8 & $-$32 40 19 & 17.31 && C && 0.42   &   \\
&&&&&&&&\\
\hline
\end{tabular}
%\end{center}
%\vspace{5mm}
\end{table*}

\begin{table*}[t]
%\begin{center}
\setcounter{table}{3}
\caption{$-$ continued}
  \label{opt3}
\begin{tabular}{lllrrcrcc}
\hline
&&&&&&&&\\

 Radio   & RA$_{J2000}$ & DEC$_{J2000}$ & S$_{22cm}$ & S$_{13cm}$&
 $\alpha^{22}_{13}$ & logP$_{22cm}$ & Radio Type   & Cluster \\

	    &	$^h$ \hspace{1mm} $^m$ \hspace{1mm} $^s$	   &
	    $^\circ$ \hspace{1mm} $^\prime$ \hspace{1mm} $^{\prime\prime}$		   
	    &  mJy	& mJy  & &W Hz$^{-1}$
	     & &     \\

 Optical  & RA$_{J2000}$ & DEC$_{J2000}$ & $b_j$	& &Cat.
 & $B_j$ & $R$  &  v$_r$ \\
	    &	$^h$ \hspace{1mm} $^m$ \hspace{1mm} $^s$	   &
	    $^\circ$ \hspace{1mm} $^\prime$ \hspace{1mm} $^{\prime\prime}$
	    &&&&&&km s$^{-1}$ \\

&&&&&&&&\\
\hline
&&&&&&&&\\
J1352$-$3314 & 13 52 08.0 & $-$33 14 45 & 20.42 &&   &&   res. &  \\
             & 13 52 07.9 & $-$33 14 52 & 18.19 && C &&  3.36$^c$ &  \\ 
             & 13 52 07.5 & $-$33 14 42 & 18.50 && C &&  3.59$^c$ &  \\  
&&&&&&&&\\
J1352$-$3315 & 13 52 09.1 & $-$33 15 26 & 26.16 &&   &&   res. &  \\
	     & 13 52 09.5 & $-$33 15 33 & 20.00 && C &&  3.93$^c$ &  \\
&&&&&&&&\\
J1352$-$3243a& 13 52 16.5 & $-$32 43 38 &  1.24 &&   && unres. & A3575 \\
	     & 13 52 16.5 & $-$32 43 38 & 17.23 && C &&  0.10  &  \\
&&&&&&&&\\
J1352$-$3306 & 13 52 33.1 & $-$33 06 01 &  6.52 & 2.95& 1.46   && unres. &  \\
             & 13 52 33.1 & $-$33 06 00 & 22.44 && C &&  0.57  &  \\
&&&&&&&&\\
J1352$-$3302 & 13 52 43.5 & $-$33 02 29 &  8.13 & 8.92&$-$0.17 && unres. &  \\
             & 13 52 43.5 & $-$33 02 29 & 18.84 && C &&  0.06  &  \\
&&&&&&&&\\
J1352$-$3251 & 13 52 50.5 & $-$32 51 24 &232.18 &145.41&0.86&& unres. &  \\
	     & 13 52 50.5 & $-$32 51 24 & 21.86 && C &&        0.10   &  \\
&&&&&&&&\\
J1353$-$3300 & 13 53 45.5 & $-$33 00 01 &  4.18 &&   && unres. &   \\
	     & 13 53 45.4 & $-$33 00 03 & 18.74 && C && 1.59   &   \\
&&&&&&&&\\
\hline
\end{tabular}
%\end{center}
\vspace{1mm}

Notes to Table 4.

$^a$ The optical counterpart is most likely a star.

$^b$ The optical candidate falls within the radio isocontours of 
the double structure. 

$^c$ These identifications have $R~>~3$, however, due to the extent
of the radio emission and/or of the optical counterpart, we can consider 
them reliable since the optical counterpart falls within the radio contours.

\end{table*}

\clearpage

\section{Radio properties of the A3571 cluster complex}

\subsection{The radio galaxies. Starburst or weak AGN?}

From the optical identification procedure discussed in Section 4.2,
only six radio sources are found to be associated with
galaxies in the Shapley Concentration. We believe this number
is underestimated due to the poor spectroscopic information available 
in this region.
Inspection of Table \ref{opt} shows that nine radio 
sources are associated with galaxies brighter than b$_J \le 18.5$ without 
redshift information. They are located
in A3571 and in A3575. It is most likely that at least a fraction of them 
belongs to the A3571 complex, and in the following we will
refer to these objects as ``candidate radio galaxies'' of the A3571
cluster complex.
We note that no radio sources are found to be associated (or possibly 
associated) with galaxies in the central region of A3572.

All the optical counterparts (with and without redshift information) are 
early-type galaxies, with the exception of J1747$-$3301 and J1349$-$3243, 
both associated with a disk galaxy.
The radio galaxies in the A3571 region are faint and unresolved by our
observations. Their radio power is in the range 
21.20 $ \le \log P_{22cm}$ (W Hz$^{-1}$) $ \le$ 22.35, i.e.  
much lower than the average transition value
between FRI and FRII radio galaxies (Fanaroff \& Riley 1974), 
i.e. $\log P_{20}$ (W Hz$^{-1}$) $\simeq$ 24.5.
Similar properties are shared by all the candidate radio galaxies. 
If we assume that they are located at the average distance of the A3571 
complex (z=0.039), only two of them have logP$_{22}$ (W Hz$^{-1}$)$ > 22$. 

It is believed that for logP$_{20}$ (W Hz$^{-1}$) $ < 22$, 
starburst becomes the dominant mechanism responsible for the radio
emission (see for instance Dwarakanath \& Owen 1999). 
Unfortunately no spectroscopy is available for the candidates in the A3571 
complex, therefore it is
not possible to establish whether they are indeed starburst galaxies or
weak active galactic nuclei, possibly driven by an ADAF nucleus
(Narayan \& Yi 1995).
Moreover, only a very small fraction of the A3571 radio galaxies and
candidates are included in the list of the 13 cm detections (see Table 
\ref{opt}), and no hint to discriminate between starburst and weak AGN can be
obtained from the higher resolution 13 cm images. 
We suggest that the A3571 complex radio galaxies and candidates
with logP$_{22}$ (W Hz$^{-1}$)~$ < 21.78$ (see Section 5.2 for radio
power division between AGN and starburst), nine in total, are starburst 
radio galaxies.

In the following, we present the images of the most interesting sources
detected in the A3571 region.\\ 

\begin{figure} [h]
%\vspace{-1cm}
\resizebox{\hsize}{!}
{\includegraphics{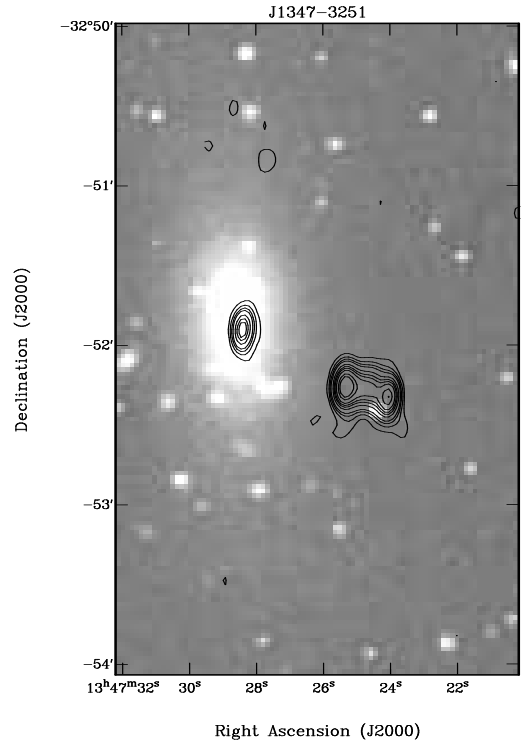}}
\caption{22 cm radio image of J1347$-$3251 overlaid on the DSS optical frame.
The contours are -0.30, 0.30, 0.60, 0.90, 1.20, 1.90, 2.50, 3.00, 3.70, 5.00,
6.00 mJy. The restoring beam is 
$11.24^{\prime\prime} \times 5.27^{\prime\prime}$, in p.a. $-7^{\circ}$.
The background double source J1347$-$3252 is also visible.}
\label{J1347-3251}
\end{figure}

\medskip
\noindent
{\it J1347$-$3251} (see Fig. \ref{J1347-3251} and Table \ref{opt}) 
is a faint radio source associated with the dominant 
cD galaxy MGC05$-$33$-$002. 
From our image, it is clear that the extension of the radio emission visible 
on the NRAO VLA Sky Survey (NVSS, Condon et al. 1998), and apparently 
associated with the cD galaxy, is actually the blend of J1347$-$3251 
itself and the double source J1347$-$3252.\\

%% figure5
%
\begin{figure} [h]
\vspace{5mm}
\resizebox{\hsize}{!}
{\includegraphics{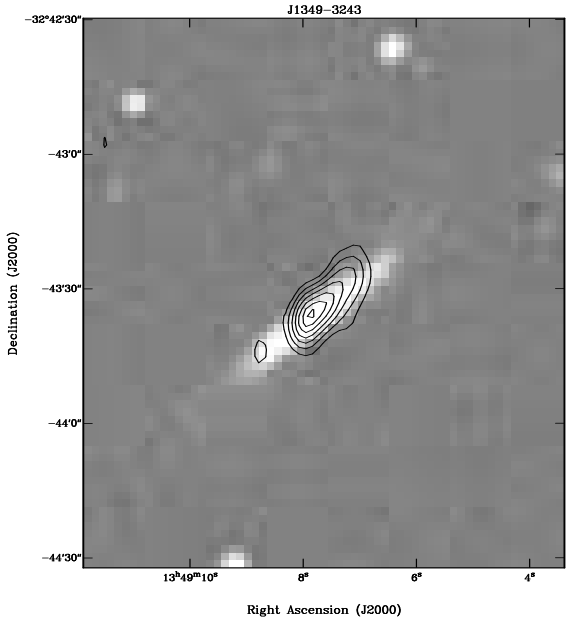}}
\caption{22 cm radio image of J1349$-$3243 overlaid on the DSS optical frame.
The contours are -0.32, 0.32, 0.48, 0.64, 0.80, 0.96, 1.10 mJy.
The restoring beam is 
$11.24^{\prime\prime} \times 5.27^{\prime\prime}$, in p.a. $-7^{\circ}$.}
\label{J1349-3243}
\end{figure}

\noindent
{\it J1349$-$3243} (see Fig. \ref{J1349-3243} and Table \ref{opt}) 
is associated 
with a disk galaxy located between the centres of A3571 and A3575.
The radio emission is elongated in the same direction of the optical 
major axis.\\

%% figure 6
%

\begin{figure} [h]
\resizebox{\hsize}{!}
{\includegraphics{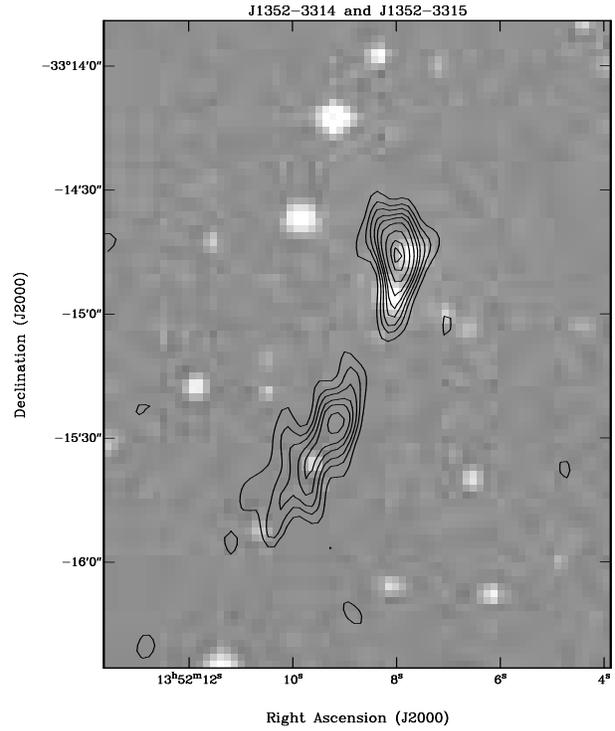}}
\caption{22 cm radio image of J1352$-$3314 and J1352$-$3315 overlaid on the DSS
optical frame. The contours are -1.10, 1.10, 1.70, 2.20, 2.80, 3.30, 4.00, 
5.00, 6.00, 6.50 mJy. The restoring beam is 
$11.24^{\prime\prime} \times 5.27^{\prime\prime}$, in p.a. $-7^{\circ}$.}
\label{J1352-3314}
\end{figure}

\medskip
In Fig. \ref{J1352-3314} the two background sources 
J1352$-$3314 and J1352$-$3315 are shown. Both sources are extended
(roughly 30$^{\prime\prime}$ and 40$^{\prime\prime}$ respectively),
and a few objects fall within the radio density contours. For this
reason the optical identification is uncertain (see Table \ref{opt}),
and we cannot exclude that they are two components of a double
source. They lie at the edge of one of the 22 cm fields, in a region
not covered by the simultaneous 13 cm observations, so no 
information is available on their radio spectrum.
%
%figure 7
%

\subsection{The radio luminosity function}

In order to investigate if cluster merger has any effect on the
formation of radio sources of nuclear origin, it is important to compare 
the radio luminosity function (RLF) for early type galaxies in 
merging environments with the local RLF derived for elliptical 
galaxies in clusters derived by Ledlow \& Owen (1996, hereinafter LO96).

We carried out this study for the A3571 complex. 
The limited number of radio galaxies found from our
survey, combined with the poor spectroscopic information available
in this region of the Shapley Concentration, does not allow a 
proper determination of the RLF, therefore in the following we will 
compare the total number of radio galaxies expected on the basis of the 
RLF derived by LO96, and the number of detections found from our survey. 
For a proper comparison, we restricted our analysis to those radio galaxies
with logP$_{22}$ (W Hz$^{-1}~) \ge$ 21.78 and with optical counterpart 
brighter than b$_J$ = 17.70 (same limits as in LO96, scaled for the proper 
cosmology and photometric band, see also Venturi et al. 2000).

A number of assumptions were made in order to compensate for the lack of 
complete spectroscopy in this region, therefore we are
aware that the following results should be considered only indicative.
In particular, we assumed that 78\%
of the 116 galaxies with b$_J \le$ 17.70 in this region, 
i.e. 90, actually belong to the A3571 complex.
This fraction was derived for the A3528 and A3558 complexes 
on the basis of a large number of spectra (Baldi, Bardelli \& Zucca 2001), 
and we believe that assuming the same percentage also for this 
nearby cluster complex is a reasonable choice.

\noindent
From Table \ref{opt} we can see that only 3 of the 6 radio
galaxies with redshift information have logP$_{22}$ (W Hz$^{-1})~ \ge$ 21.78. 
Of the remaining radio sources associated with galaxies brighter than 
b$_J$ = 17.70 and with no redshift information,
only one has logP$_{22} \ge$ 21.78 (J1349$-$3243) if placed at the
distance of the Shapley Concentration (we assumed z=0.039).

To summarise, the fraction of elliptical galaxies with b$_J \le 17.70$ and
logP$_{22} \ge$ 21.78 is 4/90 (4.4\%), to be compared to the 
12 expected ($\sim$ 14\%) on the basis of LO96. 
We point out that our estimate should be considered a lower limit, since we 
implicitly assumed that all 90 galaxies are early type. 
On the other hand, to obtain the same percentage as in LO96, only 
$\sim$ 30\% of the optical galaxies should be early type, a fraction
which seems far too low (see Fig. 5 in Baldi, Bardelli \& Zucca 2001).
In conclusion, there is an indication that the number of radio galaxies 
detected over the whole A3571 cluster complex is lower than expected on 
the basis of the RLF derived in LO96 for cluster ellipticals.

The situation changes considerably if we limit our analysis to the
cluster A3571, since all 4 radio galaxies used in the estimate of the 
RLF are located within 0.5 Abell radius (R$_A$) from the centre of A3571. 
Using the same arguments illustrated above, the upper limit to the number 
of elliptical galaxies belonging to A3571 with b$_J \le 17.70$ is 39.
The expected number of radio galaxies on the basis of LO96 is 5,
in good agreement with our result. Table \ref{rlf} summarises the results,
and shows the number of radio galaxies expected on the basis of LO96 and
the number of detections found in the whole chain (first line) and
in A3571 only (second line).

\begin{table}[ht]
%\begin{center}
\caption{Radio Luminosity Function}
\label{rlf}
\begin{tabular}{crc}
\hline
&&\\
Region & N (LO96) & N \\
        & expected & found \\
&&\\
\hline
&&\\
Whole Chain & 12 & 4 \\
A3571       & 5  & 4 \\ 
&&\\
\hline
&&\\
\end{tabular}
%\end{center}
\end{table}

\section{Radio and Optical Source Counts}

We derived the radio and source counts in the A3571 complex,
in order to check if the optical overdensity in this
region of the Shapley supercluster reflects into a higher number
of radio sources with respect to the background radio source
counts (Prandoni et al. 2001), and for comparison with the
radio counts derived for the A3558 and A3528 cluster complexes 
(Venturi et al. 2000 and 2001 respectively). 

%\subsection{Radio Source Counts} \label{rsc}

Owing to the primary beam attenuation, the sensitivity in our final image 
drops at the border of the mosaic, so the 22 cm sample discussed
in Section 4.1 is not complete to the flux limit of 0.65 mJy.
For this reason we restricted our analysis to the inner part of
the mosaic. In particular, we considered a region of 1.26 deg$^2$,
where the radio sample is complete down to the flux density of 1.30 mJy
and includes 67 radio sources.

\begin{figure} [t]
\resizebox{\hsize}{!}{\includegraphics{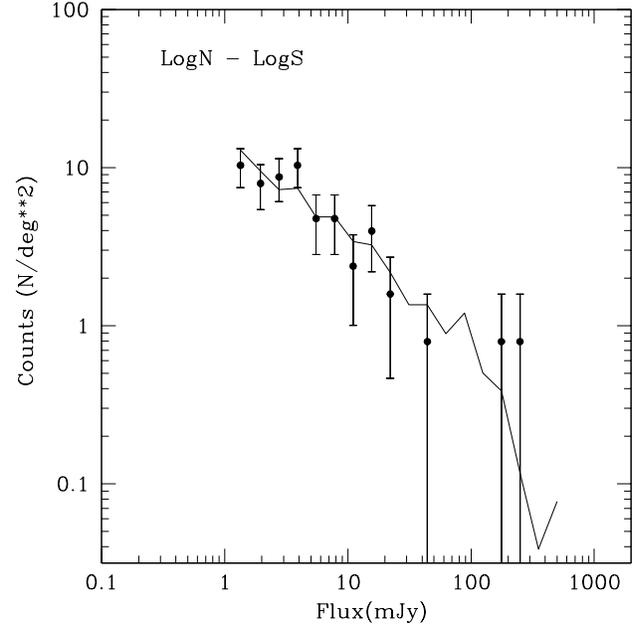}}
\caption{LogN$-$LogS. Dots refer to counts in the A3571 complex; the solid line
represents the background distribution (Prandoni et al. 2001).}
\label{cont}
\end{figure}

The LogN$-$LogS for the A3571 complex, computed in the flux density range
1.30$-$593.74 mJy, is shown in Fig. \ref{cont}. The errors are poissonian. 
We note that the first flux density bin, chosen in the range 1.16 - 1.64 mJy
for comparison with the background counts, is
incomplete. The solid line in Fig. \ref{cont} represents the normalized 
LogN$-$LogS for the background radio sources at 22 cm (Prandoni et al. 2001) 
derived over a sky region covering 25.82 deg$^2$.
% and counting a total of 1591 radio sources with $S_{22cm}$ $=$ 1.16 mJy.

% figure 8
%

Fig. \ref{cont} suggests that the two distributions have 
very similar shapes. We note that even the point in the
first bin is in good agreement with the background counts, 
despite the fact that it is incomplete.

We quantitatively estimated the similarity between the source counts in the
A3571 complex and the background counts applying a Kolmogorov-Smirnov (KS) test
to the two distributions. We found that the probability that they are the
same distribution is $p~=~$ 27\%. This percentage increases to 85\% if we
restrict the comparison to flux density $S_{22cm} \leq$ 74.22 mJy, i.e., 
dropping the high flux density bins with very few sources.
We conclude that the two distributions are statistically indistiguishable.
We found 59 radio sources in the flux range 1.64-593.74 mJy
(we exclude the incomplete flux density bins). This number is fully
consistent with what expected from the background counts, i.e. 
61$\pm$6.

The present result is in agreement with the outcome of our statistical 
analysis in the A3528 and A3558 cluster complexes, whose radio source counts 
are also consistent with the background (Venturi et al. 2000 and 2001).

%\subsection{Optical Source Counts}
%
\medskip
For comparison, we estimated the overdensity of the optical galaxies
brighter than m$_b$ = 19.5, with respect to the local background density.
We note that for the A3571 region, m$_b$ = 19.5 becomes m$_b$ = 19.3 
after correcting for the absorption, which is on average $\sim$ 0.20.

%The number of background galaxies for square degree 
%within the magnitude limit m$_b$ = 19.5, can be obtained from
%Metcalfe et al. (1994). For the A3571 region we find that the
%density of the optical background is   
%N($gal_{bck}) \simeq ~214$ gal deg$^{-2}$.

According to estimates of the optical background, made by Metcalfe et al. 
(1994) and by Bardelli et al. (1998) in a region of the Shapley 
Concentration without clusters, the projected overdensity of galaxies 
$\mathcal{O}$ with m$_b < 19.3$ in the A3571 complex is the range  
1.39 $\leq \mathcal{O_B} \leq$ 1.78, depending on the extension of
the region considered. This indicates that the 
number of galaxies in the A3571 complex is on average 50\% above the number 
expected on the basis of the background optical counts.

%the number of background galaxies 
%within the same optical magnitude limit is in the range
%$248 \leq N(gal_{bck})$ (gal deg$^{-2}$) $\leq 274$. We note that 
%these numbers include the contribution of the supercluster, therefore
%they are expected to be higher than the estimate made by Metcalfe et al.
%(1994).
%
%The density of galaxies with m$_b~<~19.3$ in an area of 5.03 deg$^2$ around 
%the A3571 chain region is N($gal_{obs}$) $=$ 306 gal deg$^{-2}$.
%
%The optical overdensity $\mathcal{O} = N(gal_{obs}){N(gal_{bck})}^{-1}$ 
%is therefore
%$\mathcal{O_M} =  1.43$ if we consider the Metcalfe background, and 
%and 1.12 $\leq \mathcal{O_B} \leq$ 1.23 for the Bardelli background.
%
%This ratio considerably increases if we limit our analysis to the 
%1.47 deg$^2$ occupied only by the A3571 complex, i.e. approximately
%the region covered by our radio survey. In this case the number
%of galaxies with m$_b~<~$ 19.3 is 
%N$(gal_{obs}) = $ 381 gal deg$^{-2}$, which leads to an overdensity
%of  
%$\mathcal{O_M} = $ 1.78 and 1.39 $\leq \mathcal{O_B} \leq$ 1.54 
%using the Metcalfe et al. (1994) background and 
%the Bardelli et al. (1998) background respectively. 

If we assume that the probability of radio emission is the same for a
field galaxy and for a Shapley galaxy, we would expect a similar excess 
also in the radio counts, i.e. 92 radio sources for an excess ratio of 1.5. 
This excess would be detectable at the 2.5$\sigma$ level, where $\sigma = 
(N_{est} - N_{bck}) \times (N_{est} + N_{bck})^{-\frac{1}{2}}$.
In particular, $N_{est}$ = estimated number of radio sources and 
$N_{bck}$ $=$ 61, as seen above.

This is in contrast with the result obtained in the radio band,
where we showed that the difference in the radio source counts and in the 
background counts is statistically negligible. This result is visible also 
from Fig. \ref{sovr}, which shows that the radio sources are fairly
uniformly distributed over the whole complex, regardless of the 
underlying optical density.

%%%%% 

\section{Discussion and Conclusions}

This work is part of a larger project whose aim is to
study the effects of cluster mergers at radio wavelengths in
the central region of the Shapley Concentration. The 
results obtained for the two major merging complexes have already
been presented and discussed (Venturi et al. 2000 and 2001). 
Here we report on ATCA 22 cm observations of the third merging complex 
in the Shapley Concentration, formed by the three clusters A3571, 
A3572 and A3575.
In particular in this paper we have focussed on the possible influence
of cluster mergers on the statistical properties of the radio galaxies.
We can briefly summarise our results as follows.

\medskip
\noindent 
{\it (a)} Six radio galaxies from our 22 cm radio sample were identified
with galaxies in the in the A3571 chain.
Other nine, associated with galaxies brighter than b$_j \le 18.5$ 
and with no measured redshift, are possible Shapley member candidates.

\noindent
{\it (b)} All Shapley radio galaxies and candidates are weak radio sources,
with radio powers in the range $21.2 < $ logP$_{22}$ (W Hz$^{-1}$) $ < 22.6$. 
They are located in A3571 and in A3575 (none in A3572).

\noindent
{\it (c)} The estimate of the RLF based on the analysis by LO96 
points towards a lack of radio galaxies in the A3571 complex when
we normalise the number of radio galaxies to the whole chain.
However, if we take into account that all the radio galaxies
considered in the RLF are located within 0.5 R$_A$ from the centre of A3571, 
the number of detections is in very good agreement with the expectations.

\noindent
{\it (d)} The radio source counts in this region are dominated
by the background counts, despite an optical overdensity of $\sim 1.5$.

\medskip
Our results suggest a dual character of the A3571 cluster complex,
where the A3571 cluster alone shows different properties than
the chain as a whole, as clear from points {\it (b)} and {\it (c)} above.
The same dual properties are evident also from observations at other 
wavelengths.

The cluster gas in A3571 is very hot (T =7.6 keV) and luminous 
(L$_{X,bol}~\sim~4\times 10^{44}$ erg s$^{-1}$, Ettori et al. 1997).
With the observed temperature we can estimate the expected 
galaxy velocity dispersion, according to the following formula 
(Lubin \& Bahcall, 1993):\\

\hspace{-5mm}$
\sigma_v = 10^{2.52\pm0.07}(kT)^{0.6\pm0.11}$\\ 

\noindent
which gives a value $\sigma_v \sim 1118$ km s$^{-1}$. This is 
in good agreement with $\sigma_v \sim 1022$ km s$^{-1}$, obtained by 
Quintana \& De Souza (1993) on the basis of redshift measurements, and
suggests that A3571 is relaxed, as proposed by Nevalainen et al. 
(2000). Another piece of evidence in favour of virialisation is 
the presence of a cooling flow in its innermost region (Peres et al.
2000), which requires that the cluster has been in equilibrium
at least since 4--6 $\times 10^9$ yrs.

Conversely, the redshift survey of Quintana et al. (1997), together with
the presence of the giant cD galaxy at the centre of A3571 
(MGC05$-$33$-$002) suggest that the whole chain is not relaxed and that 
a recent merger may be responsible for the formation of the central cD galaxy.

Inspection of the optical isodensity contours given in Fig. \ref{contour} 
shows a major optical overdensity in A3571, while the other two clusters
are not as well defined (especially A3575). In the light of the optical
distribution and of the results summarised above, it is very important
to understand whether A3575 and A3572 are the debris of a merger
event, or if they are smaller entities yet to interact with the 
massive A3571.

\medskip
The results obtained from our 22 cm survey may help in clarifying the 
situation.

It has been argued that a connection exists between cluster mergers
and a high number of starburst radio galaxies. In particular
Owen et al. (1999), in a comparative study of the two clusters A2125 and 
A2645, suggested that merging could be responsible both for the high 
fraction of blue galaxies and of radio AGN in A2125, compared to the
relaxed A2645. 

The results presented in this paper are not conclusive in this
respect, mainly because of the lack of complete spectroscopy in this
region of the sky, however there is indication that a considerable
fraction of candidate starburst galaxies is located in A3571. 
In particular, we found 9 galaxies brighter than b$_j \le 18.5$
(including confirmed members and candidates)
characterised by low radio power, i.e. logP$_{22}$ (W Hz$^{-1}$) $\le 21.78$,
whose radio emission could be driven by star formation.
These numbers should be compared to the 12 - 13
starburst galaxies detected in the richness 4 distant cluster A2125
by Owen et al. (1999). It is crucial that redshift and spectral information
becomes available for the objects in the A3571 complex, to confirm that we 
are indeed observing a high fraction of starburst galaxies. 
\noindent
We note that this result is considerably different from what we
obtained for the A3528 and A3558 complexes, where no hint of
such excess was revealed (Venturi et al. 2000 and 2001) at similar 
radio power limits.

We propose that A3571 is the result of a recent merger event, at
the very last stage of its evolution, 
and that the three main cluster complexes in the central region of 
the Shapley Concentration are part of an evolutionary sequence.
In particular, the wealth of information available from radio to
optical (both photometry and spectroscopy), up to X--ray energies, 
suggest the following scenario: 

\noindent
{\it (i)} the A3528 cluster complex is at the very beginning of a merger
event, where the two merging entities have just started ``to feel
each other''. The gradients in the temperature distribution of the 
intracluster gas delineate the merging front (Schindler 1996). The 
radio luminosity function of elliptical galaxies is in good agreement with 
that of ellipticals not located in merging clusters (LO96), and  
no sign of starburst emission, possibly induced by merger shocks, is 
detected (Venturi et al. 2001). We suggested that the pre-merging
stage had not yet had time to affect the radio emission properties 
of the cluster galaxies in the complex.

\noindent
{\it (ii)} the A3558 complex is thought to be an advanced merger,
where two clusters of similar mass have already undergone the first 
core--core encounter. The amazing similarity
between the galaxy distribution (Bardelli et al. 1998) and the gas density 
distribution (Ettori et al. 1997; Kull \& B\"ohringer 1999)
provides further evidence of the ongoing process. In the radio
band it was found (Venturi et al. 2000) that this complex shows a 
significant deficit of radio galaxies in comparison with the radio 
luminosity function of LO96, suggesting that the
major encounter switched off the nuclear radio emission and temporarily
inhibited its formation. No radio excess of starburst origin was
detected in the shock region, however data from a deeper
survey in the same region are being analysed (Venturi et al. in prep.).

\noindent
{\it (iii)} We suggest that the A3571 complex represents
the final stage of a merger event, where A3571 itself is the 
resulting cluster after virialization of the merger. The distribution
of the gas is already relaxed, as well as the galaxy distribution in
A3571, while the outer edge of the galaxy distribution is still
unrelaxed. The radio properties reflect the different dynamical
stage of the central relaxed region of the complex (the cluster
A3571) and the active dynamics of the outskirts. The location
of the radio galaxies in A3571 suggests that they had time
to develop a nuclear source after the active merging ceased.

We note that the distribution of the optical galaxies and of the 
X--ray emitting gas in these three cluster complexes is remarkably
similar to the various stages of the cluster--cluster collision
recently modelled by Ricker \& Sarazin (2001). Their simulations
for a frontal merger of two clusters with mass ratio M$_1$/M$_2$ = 1
show an A3528--like situation for t=0, which evolves into an A3558
scenario at t $\sim$ 5 Gyr, to end up in the A3571 case after 
$\sim$ 9 Gyr.

\begin{acknowledgements}
SB acknowledges support from the ASI grant ASI-I/R/037/01.
The authors wish to thank R. Stathakis for his help in the
compilation of the optical catalogue.
This work has made use of the NASA Extragalactic Database NED.
The Australia Telescope Compact Array is operated by the
Australia Telescope National Facility.
\end{acknowledgements}


\begin{thebibliography}{}

\bibitem{} Abell G.O., Corwin H.G., Olowin R.P., 1989, A\&AS, 70, 1 (ACO)
\bibitem{} Baldi A., Bardelli S., Zucca E., 2001, MNRAS 324, 509
\bibitem{} Bardelli S., Zucca E., Malizia A., Zamorani G., Vettolani G.,
Scaramella R., 1996, A\&A 305, 435
\bibitem{} Bardelli S., Zucca E., Zamorani G., Vettolani G., Scaramella R.,
1998, MNRAS, 296, 599
\bibitem{} Brunetti G., Setti G., Feretti L., Giovannini G., 2001,
MNRAS 320, 365
\bibitem{} Buote D.A., 2001, ApJL, 533, 15
\bibitem{} Condon J.J., Cotton W.D., Greisen E.W., Yin F., 1998, AJ 115, 1693
\bibitem{} Corwin H.G., de Vaucouleurs A., de Vaucouleurs G., 1985, SGC, C, 
0000C
\bibitem{} Da Costa L.N., Nunes P.S., Pellegrini P.S., Willmer C., 
Chincarini G., Cowan J.J., 1986, Aj, 91, 6 
\bibitem{} Dressler A., 1991, ApJS, 75, 241
\bibitem{} Drinkwater M.J., Proust D., Parker Q.A., Quintana H., Slezak E.,
1999, PASA, 16, 113
\bibitem{} Dwarakanath K.S., Owen F.N., 1999, AJ 118, 625
\bibitem{} Edge A.C., Stewart G.C., Fabian A.C., Arnaud K.A., 1990, MNRAS, 245,
559
\bibitem{} Ensslin T.A., Br\"uggen M., 2001, MNRAS, in press
\bibitem{} Ettori S., Fabian A.C., White D.A., 1997, MNRAS, 289, 787
\bibitem{} Fanaroff \& Riley, 1974, MNRAS ...
\bibitem{} Feretti L., Giovannini G., 1996, in {\it Extragalactic Radio 
Sources}, IAU Symp. 175, Eds. R. Ekers, C. Fanti \& L. Padrielli,
Kluwer Academic Publ., p. 133
\bibitem{} Feretti L., Giovannini G., 2001, in {\it Cluster Mergers},
Eds. L. Feretti, I.M. Gioia \& G. Giovannini, Kluwer Academic Publ., in press
\bibitem{} Girardi M., Giuricin G., Mardirossian F., Mezzetti M., Boschin W.,
1998, ApJ 505, 74
\bibitem{} Hanami H.,  Tsuru T., Shimasaku K., Yamauchi S., Ikebe Y.,
Koyama K., 1999, ApJ 521, 90
\bibitem{} Heisler J., Tremaine S., Bahcall J.N., 1985, ApJ, 298, 8 
\bibitem{} Johnstone R.M., Naylory T., Fabian A.C., 1991, MNRAS, 248, 18
\bibitem{} Kemp S.N., Meaburn J., 1991, MNRAS, 251, 10
\bibitem{} Kull A., B\"ohringer H., 1999, A\&A 341, 23
\bibitem{} Lahav O., Edge A.C., Fabian A.C., Putney A., 1989, MNRAS, 238, 881
\bibitem{} Lauberts A., 1982, ESOU, C, 0000L
\bibitem{} Ledlow M.J., Owen F.N., 1996, ApJ, 112, 9 (LO96)
\bibitem{} Lubin L.M., Bahcall N.A., 1993, ApJ, 415, 17
\bibitem{} Maddox S.J., Efstathiou G., Sutherland W.J., Loveday J., 1990,
MNRAS 243, 692
\bibitem{} Metcalfe N., Godwin J.G., Peach J.V., 1994, MNRAS, 267, 431
\bibitem{} Narayan R., Yi I., 1995, ApJ 452, 710 
\bibitem{} Nevalainen J., Markevitch M., Forman W., 2000, ApJ, 536, 73
\bibitem{} Nevalainen J., Kaastra J., Parmar A.N., Markevitch M., Oosterbroek
T., Colafrancesco S., Mazzotta P., 2001, A\&A 369, 459
\bibitem{} Peres C.B., Fabian A.C., Edge A.C., Allen S.W., Johnstone R.M., 
White D.A., 1998, MNRAS, 298, 416
\bibitem{} Prandoni I., Gregorini L., Parma P., de Ruiter H.R., Vettolani
G., Wieringa M.H., Ekers R.D., 2000a, A\&A Suppl. Ser. 146, 31
\bibitem{} Prandoni I., Gregorini L., Parma P., de Ruiter H.R., Vettolani
G., Wieringa M.H., Ekers R.D., 2000b, A\&A Suppl. Ser. 146, 41 
\bibitem{} Prandoni I., Gregorini L., Parma P., de Ruiter H.R., Vettolani
G., Wieringa M.H., Ekers R.D., 2001, A\&A 365, 392
\bibitem{} Quintana H., de Souza R., 1993, A\&AS, 101, 475
\bibitem{} Quintana H., Melnick J., Proust D., Infante L., 1997, A\&AS, 
125, 247 
\bibitem{} Ricker P.M., Sarazin C.L., 2001, ApJ in press, astro-ph/0107210
\bibitem{} Sarazin C.L., 2000, in {\it Constructing the Universe with
Clusters of Galaxies}, Eds. Durrett F. \& Gerbal D., electronic
proc. http://www.iap.fr/Conferences/Colloque/coll2000/contributions
\bibitem{} Sault R.J., Teuben P.J., Wright M.C.H., 1995, in {\it Astronomical
Data Analysis Software and Systems. IV} eds. Shaw R., Peyne H.E. \& 
Hayes J.J.E., Astronomic Society of the Pacific Conference Series 77, 433
\bibitem{} Schindler S., 1996, MNRAS 280, 309
\bibitem{} Unewisse A.M., Hunstead D.W., Pietrzynski B., 1993, Pubbl.
Astron. Soc. Austr. 10, 229
\bibitem{} Venturi T., Bardelli S., Morganti R., Hunstead R.W., 2000, MNRAS,
314, 594
\bibitem{} Venturi T., Bardelli S., Zambelli G., Morganti R., Hunstead R.W., 
2001, MNRAS 324, 1131
\bibitem{} Vorontsov-Velyaminov B.A., Arhipova V.P., 1974, in 
{\it Morphological Catalog of Galaxies} (MCG), V, Moscow State University
\bibitem{} Yentis D.J., Cruddace R.G., Gursky H.,  Stuart B.V., Wallin J.F., 
MacGillivray H.T., Collins C.A., 1992, in {\it Digitised Optical Sky Surveys},
Editors, H.T. MacGillivray, E.B. Thomson; Publisher, Kluwer Academic 
Publishers, p. 67
\bibitem{} Zucca E., Zamorani G., Scaramella R., Vettolani G., 1993, ApJ, 407,
470


\end{thebibliography}
\end{document}